\documentclass[12pt]{article}
\def\be{\begin{equation}}
\def\ee{\end{equation}}
\def\bea{\begin{eqnarray}}
\def\eea{\end{eqnarray}}
\usepackage{graphicx}

\catcode`\@=11
\def\lsim{\mathrel{\mathpalette\@versim<}}
\def\gsim{\mathrel{\mathpalette\@versim>}}
\def\@versim#1#2{\vcenter{\offinterlineskip
\ialign{$\m@th#1\hfil##\hfil$\crcr#2\crcr\sim\crcr } }}
\catcode`\@=12
\usepackage{axodraw}

\catcode`@=12
\begin{document}
\thispagestyle{empty}
\begin{flushright}
UCRHEP-T507\\
July 2011\
\end{flushright}
\vspace{0.3in}
\begin{center}
{\LARGE \bf Dark Vector-Gauge-Boson Model\\}
\vspace{1.5in}
{\bf Subhaditya Bhattacharya$^1$, J. Lorenzo Diaz-Cruz$^2$,\\ Ernest Ma$^1$, 
and Daniel Wegman$^1$\\}
\vspace{0.2in}
{\sl $^1$ Department of Physics and Astronomy, University of California,\\ 
Riverside, California 92521, USA\\
$^2$ Facultad de Ciencias Fisico-Matematicas, \\
Benemerita Universidad Autonoma de Puebla, Puebla, Mexico \\}
\end{center}
\vspace{1.2in}
\begin{abstract}\
A model based on $SU(3)_C \times SU(2)_L \times U(1)_Y \times SU(2)_N$ has 
recently been proposed, where the $SU(2)_N$ vector gauge bosons are neutral, 
so that a vector dark-matter candidate is possible and constrained by data 
to be less than about 1 TeV.  We explore further implications of this model, 
including a detailed study of its Higgs sector.  We improve on its dark-matter 
phenomenology, as well as its discovery reach at the LHC 
(Large Hadron Collider).
\end{abstract}

\newpage
\baselineskip 24pt

\section{Introduction}

The nature of dark matter~\cite{bhs05} is under intense study.  Whereas 
most assume that it is either a fermion or a scalar or a combination of 
both~\cite{cmwy07}, the notion that it could be a vector boson just as 
well has also been proposed. In a theory of universal compact extra 
dimensions, the first Kaluza-Klein excitation of the standard-model $U(1)$ 
gauge boson $B$ is such a candidate~\cite{st03}. The $T-$odd counterpart of 
$B$ in little Higgs models is another candidate~\cite{hm05}. Non-Abelian 
vector bosons from a hidden sector may also be considered~\cite{h09}. 
All of the above involve ``exotic'' physics.

Recently, it was realized~\cite{dm11} that an existing conventional 
model~\cite{lr86} based on superstring-inspired $E_6$ has exactly the 
ingredients which allow 
it to become a model of vector-boson dark matter, where the vector boson  
itself ($X$) comes from an $SU(2)_N$ gauge extension of the Standard Model.
In Sec.~2 we list all the necessary particles of this (nonsupersymmetric) 
model.  In Sec.~3 we discuss in detail the complete Higgs potential 
and its minimization.  In Sec.~4 we obtain the masses of all the gauge and 
Higgs bosons.  In Sec.~5 we compute the annihilation cross section of the 
dark-mmatter vector boson $X$.  In Sec.~6 we study the constraints from 
dark-matter direct-search experiments.  In Sec.~7 we consider some possible 
signals at the Large Hadron Collider (LHC).  In Sec.~8 there are some 
concluding 
remarks.

\section{Particle content}

Under $SU(3)_C \times SU(2)_L \times U(1)_Y \times SU(2)_N \times S$, 
where $Q = T_{3L} + Y$ is the electric charge and $L = S + T_{3N}$ is the 
generalized lepton number, the fermions of this nonsupersymmetric model 
are given by~\cite{dm11}
\begin{eqnarray}
&& \pmatrix{u \cr d} \sim (3,2,1/6,1;0), ~~~ u^c \sim (3^*,1,-2/3,1;0), \\ 
&& (h^c,d^c) \sim (3^*,1,1/3,2;-1/2), ~~~ h \sim (3,1,-1/3,1;1), \\ 
&& \pmatrix{N & \nu \cr E & e} \sim (1,2,-1/2,2;1/2), ~~~ \pmatrix{E^c \cr N^c} 
\sim (1,2,1/2,1;0), \\ 
&& e^c \sim (1,1,1,1;-1), ~~~ (\nu^c,n^c) \sim (1,1,0,2;-1/2),
\end{eqnarray}
where all fields are left-handed.  The $SU(2)_L$ doublet assignments 
are vertical with $T_{3L} = \pm 1/2$ for the upper (lower) entries. 
The $SU(2)_N$ doublet assignments are horizontal with $T_{3N} = \pm 1/2$ 
for the right (left) entries.  There are three copies of the above to 
accommodate the known three generations of quarks and leptons, together 
with their exotic counterparts.  It is easy to check that all gauge 
anomalies are canceled. The extra global U(1) symmetry $S$ is imposed so 
that $(-1)^L$, where $L = S + T_{3N}$, is conserved, even though $SU(2)_N$ 
is completely broken.

The Higgs sector consists of one bidoublet, two doublets, and one triplet:
\begin{eqnarray}
&& \pmatrix{\phi_1^0 & \phi_3^0 \cr \phi_1^- & \phi_3^-} \sim (1,2,-1/2,2;1/2), 
~~~ \pmatrix{\phi_2^+ \cr \phi_2^0} \sim (1,2,1/2,1;0), \\ 
&& (\chi_1^0,\chi_2^0) \sim (1,1,0,2;-1/2), ~~~ \pmatrix{\Delta^0_2/\sqrt{2} 
& \Delta^0_3 \cr \Delta^0_1 & -\Delta^0_2/\sqrt{2}} \sim (1,1,0,3;1).
\end{eqnarray}
The allowed Yukawa couplings are thus
\begin{eqnarray}
&& (d \phi_1^0 - u \phi_1^-) d^c - (d \phi_3^0 - u \phi_3^-) h^c, ~~~ 
(u \phi_2^0 - d \phi_2^+) u^c, ~~~ (h^c \chi_2^0 - d^c \chi_1^0) h, \\ 
&& (N \phi_3^- - \nu \phi_1^- - E \phi_3^0 + e \phi_1^0) e^c, ~~~ 
(E \phi_2^+ - N \phi_2^0) n^c - (e \phi_2^+ - \nu \phi_2^0) \nu^c, \\ 
&& (E E^c - N N^c) \chi_2^0 - (e E^c - \nu N^c) \chi_1^0, ~~~ 
n^c n^c \Delta^0_1 + (n^c \nu^c + \nu^c n^c) \Delta^0_2/\sqrt{2} - 
\nu^c \nu^c \Delta^0_3.
\end{eqnarray}
There are five nonzero vacuum expectation values: $\langle \phi_1^0 
\rangle = v_1$, $\langle \phi_2^0 \rangle = v_2$, $\langle \Delta_1^0 
\rangle = u_1$, and $\langle \chi_2^0 \rangle = u_2$,  corresponding to 
scalar fields with $L=0$, as well as $\langle \Delta_3^0 \rangle = u_3$, 
which breaks $L$ to $(-1)^L$.   Thus $m_d, m_e$ come from 
$v_1$, and $m_u, m_{\nu \nu^c} (= -m_{N n^c})$ come from $v_2$, whereas 
$m_h, m_E (= -m_{N N^c})$ come from $u_2$, and $n^c$, $\nu^c$ obtain 
Majorana masses from $u_1$ and $u_3$. The scalar fields $\phi_3^{0,-}$ 
and $\Delta_2^0$ have $L=1$, whereas $\chi_1^0$ has $L=-1$ and 
$\Delta_3^0$ has $L=2$. 

There are five neutral fermions per family.  Two have odd $L$ parity, i.e. 
$\nu$ and $\nu^c$.  Their $2 \times 2$ mass matrix is of the usual seesaw 
form, i.e.
\begin{equation}
{\cal M}_\nu = \pmatrix{0 & m_D \cr m_D & M_3},
\end{equation}
where $m_D$ comes from $v_2$ and $M_3$ from $u_3$.  The other three have 
even $L$ parity, i.e. $N$, $N^c$, and $n^c$.  Their $3 \times 3$ mass matrix 
is given by
\begin{equation}
{\cal M}_N = \pmatrix{0 & -m_E & -m_D \cr -m_E & 0 & 0 \cr -m_D & 0 & M_1},
\end{equation}
where $m_E$ comes from $u_2$ and $M_1$ from $u_1$.  Note that without 
$M_1$, there would be a massless fermion in this sector.  Since $(-1)^L$ 
is exactly conserved, $\nu,\nu^c$ do not mix with $N,N^c,n^c$.

Even though this model is nonsupersymmetric, $R$ parity as defined in the 
usual way for supersymmetry, i.e. $R \equiv (-)^{3B+L+2j}$, still holds, 
so that the usual quarks and leptons have even $R$, whereas $h, h^c, (N,E), 
(E^c,N^c)$, and $n^c$ have odd $R$.  As for the scalars, $(\phi_1^0,\phi_1^-)$, 
$(\phi_2^+,\phi_2^0)$, $\chi_2^0$, $\Delta_1^0$, and $\Delta_3^0$ have even 
$R$, whereas $(\phi_3^0,\phi_3^-)$, $\chi_1^0$, and $\Delta_2^0$ have odd 
$R$.

\section{Higgs potential}
The Higgs potential of this model is given by
\begin{eqnarray}
V &=& \mu_1^2 Tr(\phi_{13}^\dagger \phi_{13}) + \mu_2^2 \phi_2^\dagger \phi_2 
+ \mu_\chi^2 \chi \chi^\dagger + \mu_\Delta^2 Tr(\Delta^\dagger \Delta) \nonumber 
\\ &+& (\mu_{22} \tilde{\chi} \phi_{13}^\dagger \tilde{\phi}_2 + \mu_{12} 
\chi \Delta \tilde{\chi}^\dagger + \mu_{23} \tilde{\chi} \Delta {\chi}^\dagger 
+ H.c.) +  {1 \over 2} \lambda_1 [Tr(\phi_{13}^\dagger \phi_{13})]^2 + 
{1 \over 2} \lambda_2 (\phi_2^\dagger \phi_2)^2  \nonumber \\ 
&+&  {1 \over 2} \lambda_3 Tr(\phi_{13}^\dagger \phi_{13} \phi_{13}^\dagger 
\phi_{13}) + {1 \over 2} \lambda_4 (\chi \chi^\dagger)^2  +{1 \over 2} 
\lambda_5 [Tr(\Delta^\dagger \Delta)]^2 + {1 \over 4} 
\lambda_6 Tr(\Delta^\dagger \Delta - \Delta \Delta^\dagger)^2 \nonumber \\ 
&+& f_1 \chi \phi_{13}^\dagger \phi_{13} \chi^\dagger + f_2 \chi 
\tilde{\phi}_{13}^\dagger \tilde{\phi}_{13} \chi^\dagger  + f_3 \phi_2^\dagger 
\phi_{13} \phi_{13}^\dagger \phi_2 + f_4 \phi_2^\dagger \tilde{\phi}_{13} 
\tilde{\phi}_{13}^\dagger \phi_2 + f_5 (\phi_2^\dagger \phi_2)(\chi \chi^\dagger) 
\nonumber \\ &+& f_6 (\chi \chi^\dagger) 
Tr(\Delta^\dagger \Delta) + f_7 \chi (\Delta^\dagger \Delta - \Delta 
\Delta^\dagger) \chi^\dagger + f_8 (\phi_2^\dagger \phi_2) 
Tr(\Delta^\dagger \Delta) \nonumber \\ &+& f_9 
Tr(\phi_{13}^\dagger \phi_{13}) Tr(\Delta^\dagger \Delta) + f_{10} 
Tr(\phi_{13} (\Delta^\dagger \Delta - \Delta \Delta^\dagger) \phi_{13}^\dagger),
\end{eqnarray}
where
\begin{equation}
\tilde{\phi}_2 = \pmatrix{\bar{\phi}_2^0 \cr -\phi_2^-}, ~~~ \tilde{\phi}_{13} = 
\pmatrix{\phi_3^+ & -\phi_1^+ \cr - \bar{\phi}_3^0 & \bar{\phi}_1^0}, ~~~ 
\tilde{\chi} = (\bar{\chi}_2^0, - \bar{\chi}_1^0),
\end{equation}
and the $\mu_{23}$ term breaks $L$ softly to $(-1)^L$.

The minimum of $V$ is determined by
\begin{eqnarray}
V_0 &=& \mu_1^2 v_1^2 + \mu_2^2 v_2^2 + \mu_\chi^2 u_2^2 + \mu_\Delta (u_1^2 
+ u_3^2) + 2 \mu_{22} v_1 v_2 u_2 + 2 \mu_{12} u_1 u_2^2 + 2 \mu_{23} u_3 u_2^2 
\nonumber \\ &+& {1 \over 2} \lambda_1 v_1^4 + {1 \over 2} \lambda_2 v_2^4 
+ {1 \over 2} \lambda_3 v_1^4 + {1 \over 2} \lambda_4 u_2^4 + {1 \over 2} 
\lambda_5 (u_1^2 + u_3^2)^2 + {1 \over 2} \lambda_6 (u_1^2 - u_3^2)^2 
\nonumber \\ &+& f_2 v_1^2 u_2^2 + f_4 v_1^2 v_2^2 + f_5 v_2^2 u_2^2 
+ f_6 u_2^2 (u_1^2 + u_3^2) + f_7 u_2^2 (u_3^2 - u_1^2) \nonumber \\ &+& 
f_8 v_2^2 (u_1^2 + u_3^2) + f_9 v_1^2 (u_1^2 + u_3^2) + f_{10} v_1^2 
(u_1^2 - u_3^2),
\end{eqnarray}
where
\begin{eqnarray}
0 &=& \mu_1^2 + (f_9 + f_{10}) u_1^2 + f_2 u_2^2 + (f_9 - f_{10}) u_3^2 + 
(\lambda_1 + \lambda_3) v_1^2 + f_4 v_2^2 + {\mu_{22} v_2 u_2 \over v_1}, \\ 
0 &=& \mu_2^2 + f_8 u_1^2 + f_5 u_2^2 + f_8 u_3^2 + f_4 v_1^2 + \lambda_2 v_2^2 
+ {\mu_{22} v_1 u_2 \over v_2}, \\ 
0 &=& \mu_\chi^2 + (f_6 - f_7) u_1^2 + \lambda_4 u_2^2 + (f_6 + f_7) u_3^2 
+ f_2 v_1^2 + f_5 v_2^2 + {\mu_{22} v_1 v_2 \over u_2} \nonumber \\ 
&+& 2 \mu_{12} u_1 + 2 \mu_{23} u_3, \\ 
0 &=& \mu_\Delta^2 + (\lambda_5 + \lambda_6) u_1^2 + (f_6 - f_7) u_2^2 + 
(\lambda_5 - \lambda_6) u_3^2 + (f_9 + f_{10}) v_1^2 + f_8 v_2^2 + 
{\mu_{12} u_2^2 \over u_1}, \\ 
0 &=& \mu_\Delta^2 + (\lambda_5 - \lambda_6) u_1^2 + (f_6 + f_7) u_2^2 + 
(\lambda_5 + \lambda_6) u_3^2 + (f_9 - f_{10}) v_1^2 + f_8 v_2^2 + 
{\mu_{23} u_2^2 \over u_3}. 
\end{eqnarray}

\section{Gauge and Higgs boson masses}
After the spontaneous breaking of $SU(2)_N \times SU(2)_L \times U(1)_Y$, the 
gauge bosons $X_{1,2,3}$ and $W,Z$ acquire masses as follows:
\begin{eqnarray}
&& m_W^2 = {1 \over 2} g_2^2 (v_1^2 + v_2^2), ~~~ m_{X_{1,2}}^2 = {1 \over 2} 
g_N^2 [u_2^2 + v_1^2 + 2(u_1 \mp u_3)^2], \\ 
&& m^2_{Z,X_3} = {1 \over 2} \pmatrix{ (g_1^2+g_2^2)(v_1^2+v_2^2) & -g_N 
\sqrt{g_1^2 + g_2^2} v_1^2 \cr -g_N \sqrt{g_1^2 + g_2^2} v_1^2 & 
g_N^2 [u_2^2 + v_1^2 + 4(u_1^2 + u_3^2)]}.
\end{eqnarray}
Whereas the usual gauge bosons have even $R$, two of the $SU(2)_N$ gauge 
bosons $X_{1,2}$ have odd $R$ and $X_3 (=Z')$ has even $R$.  Assuming that 
$X_1$ is lighter than $X_2$, the former becomes a good candidate for dark 
matter.  There is also $Z-Z'$ mixing in this model, given by 
$-(\sqrt{g_1^2+g_2^2}/g_N)[v_1^2/(u_2^2+4u_1^2+4u_3^2)]$.  This is constrained 
by precision electroweak data to be less than a few times $10^{-4}$.  If 
$m_{Z'} \sim 1$ TeV, then $v_1$ should be less than about 10 GeV.  Now $m_b$ 
comes from $v_1$, so this model implies that $\tan \beta = v_2/v_1$ is 
large and the Yukawa coupling of $b b^c \phi_1^0$ is enhanced.  This will 
have interesting phenomenological consequences~\cite{bdhty99}.

There are 22 scalar degrees of freedom, 6 of which become massless Goldstone 
bosons, leaving 16 physical particles.  Their masses are given below:
\begin{eqnarray}
&& m^2(\phi_3^\pm) = (f_1-f_2)u_2^2 + 2f_{10}(u_3^2-u_1^2) - \lambda_3 v_1^2 
+ (f_3-f_4) v_2^2 - \mu_{22} v_2 u_2/v_1, \\ 
&& m^2(\sin \beta \phi_1^\pm + \cos \beta \phi_2^\pm) = [f_3 - f_4 - \mu_{22} 
u_2/v_1 v_2] \sqrt{v_1^2 + v_2^2},
\end{eqnarray}
where $\tan \beta = v_2/v_1$ and the orthogonal combination $\cos \beta 
\phi_1^\pm - \sin \beta \phi_2^\pm$ is massless, corresponding to the 
longitudinal component of $W^\pm$.  The $5 \times 5$ mass-squared 
matrix spanning $(\phi_{1I}, \phi_{2I}, \chi_{2I}, \Delta_{1I}, \Delta_{3I})$ 
is given by
\begin{equation}
\pmatrix{-\mu_{22}v_2u_2/v_1 & -\mu_{22}u_2 & -\mu_{22}v_2 & 0 
& 0 \cr -\mu_{22}u_2 & -\mu_{22}v_1u_2/v_2 & -\mu_{22}v_1 & 0 & 0 \cr 
-\mu_{22}v_2 & -\mu_{22}v_1 & -\mu_{22}v_1v_2/u_2 - 4\mu_{12}u_1 - 4\mu_{23}u_3 & 
-2\mu_{12}u_2 & 2\mu_{23}u_2 \cr 0 & 0 & -2\mu_{12}u_2 & -\mu_{12}u_2^2/u_1 & 0 
\cr 0 & 0 & 2\mu_{23}u_2 & 0 & -\mu_{23}u_2^2/u_3},
\end{equation}
with two zero mass eigenvalues, spanned by the states $v_1 \phi_{1I} - v_2 
\phi_{2I}$ and $-(v_1/2) \phi_{1I} -(v_2/2) \phi_{2I} + u_2 \chi_{2I} - 2u_1 
\Delta_{1I} + 2 u_3 \Delta_{3I}$, corresponding to the longitudinal components 
of $Z$ and $Z'$.  In the $(\chi_{1I}, \Delta_{2I}, \phi_{3I})$ sector, the 
mass-squared matrix is given by
\begin{eqnarray}
&& [(f_1-f_2)v_1^2 + 2f_7(u_1^2-u_3^2)-\mu_{22}v_1v_2/u_2 - 2(\mu_{12}-\mu_{23})
(u_1-u_3)] \chi_{1I}^2 \nonumber \\ &+& 2\sqrt{2}u_2[\mu_{23}-\mu_{12} + 
f_7 (u_1+u_3)] \chi_{1I} \Delta_{2I} + 2[\mu_{22}v_2 - (f_1-f_2)v_1u_2] \chi_{1I} 
\phi_{3I} \nonumber \\
&+& [\lambda_6 (u_1+u_3)^2 - \mu_{12}u_2^2/2u_1 - \mu_{23}u_2^2/2u_3] 
\Delta_{2I}^2 - 2\sqrt{2} f_{10} v_1(u_3+u_1) \Delta_{2I} \phi_{3I} \nonumber \\ 
&+& [(f_1-f_2)u_2^2 + 2f_{10}(u_3^2-u_1^2) - \mu_{22} v_2u_2/v_1] \phi_{3I}^2,
\end{eqnarray}
with one zero mass eigenvalue, corresponding to the longitudinal component 
of $X_1$.  The mass-squared matrix of the $(\chi_{1R}, \Delta_{2R}, \phi_{3R})$ 
sector is analogously given by
\begin{eqnarray}
&& [(f_1-f_2)v_1^2 + 2f_7(u_1^2-u_3^2)-\mu_{22}v_1v_2/u_2 - 2(\mu_{12}+\mu_{23})
(u_1+u_3)] \chi_{1R}^2 \nonumber \\ &+& 2\sqrt{2}u_2[\mu_{23}+\mu_{12}
+f_7(u_3-u_1)] \chi_{1R} 
\Delta_{2R} - 2[\mu_{22}v_2 - (f_1-f_2)v_1u_2] \chi_{1R} \phi_{3R} \nonumber \\
&+& [\lambda_6 (u_1-u_3)^2 - \mu_{12}u_2^2/2u_1 - \mu_{23}u_2^2/2u_3] 
\Delta_{2R}^2 + 2\sqrt{2} f_{10} v_1(u_3-u_1) \Delta_{2R} \phi_{3R} \nonumber \\ 
&+& [(f_1-f_2)u_2^2 + 2f_{10}(u_3^2-u_1^2) - \mu_{22} v_2u_2/v_1] \phi_{3R}^2,
\end{eqnarray}
with one zero mass eigenvalue, corresponding to the longitudinal component 
of $X_2$.  The remaining 5 scalar fields $(\phi_{1R},\phi_{2R},\chi_{2R}, 
\Delta_{1R}, \Delta_{3R})$ form a mass-squared matrix
\begin{eqnarray}
&& \pmatrix{2(\lambda_1+\lambda_3) v_1^2 & 2f_4 v_1 v_2 & 2 f_2 v_1 u_2 & 
2(f_9+f_{10})v_1u_1 & 2(f_9+f_{10})v_1u_3 \cr 2f_4 v_1 v_2 & 2\lambda_2 v_2^2 
& 2 f_5 v_2 u_2 & 2 f_8 v_2 u_1 & 2 f_8 v_2 u_3 \cr 2 f_2 v_1 u_2 & 2 f_5 v_2 
u_2 & 2 \lambda_4 u_2^2 & 2(f_6-f_7)u_1u_2 & 2(f_6+f_7)u_2u_3 \cr 
2(f_9+f_{10})v_1 u_1 & 2 f_8 v_2 u_1 & 2(f_6-f_7)u_1u_2 & 2(\lambda_5+\lambda_6) 
u_1^2 & 2(\lambda_5-\lambda_6)u_1u_3 \cr 2(f_9-f_{10})v_1u_3 & 2f_8 v_2 u_3 & 
2(f_6+f_7)u_2u_3 & 2(\lambda_5-\lambda_6)u_1u_3 & 2(\lambda_5+\lambda_6)u_3^2} 
\nonumber \\ 
&+& \pmatrix{-\mu_{22}v_2u_2/v_1 & \mu_{22}u_2 & \mu_{22} v_2 & 0 & 0 \cr 
\mu_{22} u_2 & -\mu_{22}v_1u_2/v_2 & \mu_{22} v_1 & 0 & 0 \cr 
\mu_{22} v_2 & \mu_{22} v_1 & -\mu_{22} v_1 v_2/u_2 & 2\mu_{12} u_2 & 2\mu_{23} 
u_2 \cr 0 & 0 & 2\mu_{12} u_2 & -\mu_{12}u_2^2/u_1 & 0 \cr 0 & 0 & 
2 \mu_{23}u_2 & 0 & -\mu_{23}u_2^2/u_3}.
\end{eqnarray}

Consider the simplifying case of $f_7=f_{10}=0$ and $\mu_{12}=\mu_{23}$, then 
from Eqs.~(18) and (19), we find $u_1=u_3$.  The massless states of Eqs.~(25) 
and (26) are then easily identified: $u_2 \chi_{1I} + v_1 \phi_{3I}$ and 
$u_2 \chi_{1R} + 2 \sqrt{2} u_1 \Delta_{2R} - v_1 \phi_{3R}$ for the 
longitudinal components of $X_1$ and $X_2$ respectively.  Three exact 
mass eigenstates are:
\begin{eqnarray}
(\Delta_{1I} + \Delta_{3I})/\sqrt{2}: && m^2 = -\mu_{12}u_2^2/u_1, \\ 
\Delta_{2I},~(\Delta_{1R} - \Delta_{3R})/\sqrt{2}: && m^2 = 4 \lambda_6 u_1^2 
- \mu_{12}u_2^2/u_1.
\end{eqnarray}
Using the approximation $v_{1,2} << u_{1,2}$, we also have
\begin{eqnarray}
\phi_{3R},~\phi_{3I}: && m^2 = (f_1-f_2)u_2^2 - \mu_{22}u_2v_2/v_1, \\ 
(v_2 \phi_{1I} + v_1 \phi_{2I})/\sqrt{v_1^2+v_2^2}: && m^2 = -\mu_{22} u_2 
(v_1^2 + v_2^2)/v_1v_2, \\ 
(2\sqrt{2}u_1 \chi_{1R} - u_2 \Delta_{2R})/\sqrt{8 u_1^2 + u_2^2} 
: && m^2 = - \mu_{12} (8u_1 + u_2^2/u_1), \\ 
(4u_1 \chi_{2I} + u_2 \Delta_{1I} - u_2 \Delta_{3I})/ \sqrt{16 u_1^2 
+ 2 u_2^2}: && m^2 = - \mu_{12} (8u_1 + u_2^2/u_1).
\end{eqnarray}

This pattern shows that $(\phi_1^0,\phi_1^-)$ and $(\phi_2^+,\phi_2^0)$ 
behave as the conventional two Higgs doublets with the former coupling to 
$d$ quarks and and the latter to $u$ quarks.  The new feature here is 
that $(\phi_1^0,\phi_1^-)$ also interact with the $SU(2)_N$ gauge bosons. 
An interesting possibility for example is $Z' \to \phi_1^0 \bar{\phi}_1^0 \to 
(b \bar{b})(b \bar{b})$.

\section{$X_1 X_1$ annihilation}
We assume that $X_1$ is the lightest particle having odd $R$.  It is thus 
stable and a possible candidate for dark matter.  In the early Universe, 
$X_1 X_1$ will annihilate to particles of even $R$, i.e. $d \bar{d}$ through 
$h$ exchange, $e^- e^+$ through $E$ exchange, $\nu \bar{\nu}$ through $N$ 
exchange, and $\phi_1 \bar{\phi}_1$ through $\phi_3$ exchange (and direct 
interaction).  There is also the direct-channel process, such as $X_1 X_1 \to 
\phi_{1R} \to d \bar{d}$, which is suppressed by $m_d$ so it is negligible 
here.  However, the corresponding process for dark-matter direct search, 
i.e. $X_1 d \to X_1 d$ through $\phi_{1R}$ exchange, may be important 
as discussed in the next section.  Note that there is no tree-level 
contribution from $Z'$ because the only allowed triple-vector-boson 
coupling  is $X_1 X_2 Z'$ and $X_2$ is too heavy to be involved.

\begin{figure}[htb]
\begin{center}
\includegraphics[width=0.7\textwidth]{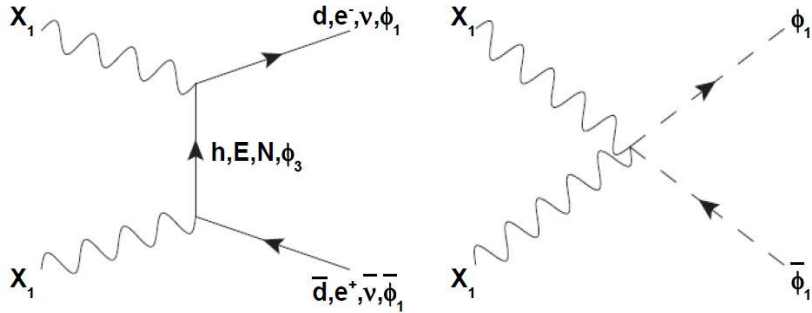}
\caption{Annihilation of $X_1 X_1$ to standard-model particles.}
\end{center}
\end{figure}
In Fig.~1 we show the various annihilation diagrams, resulting in the 
nonrelativistic cross section $\times$ relative velocity given by
\begin{eqnarray}
\sigma v_{rel} &=& {g_N^4 m_X^2 \over 72 \pi} \left[ \sum_h {3 \over (m_h^2 
+ m_X^2)^2} + \sum_E {2 \over (m_E^2 + m_X^2)^2}  \right. \nonumber \\ 
&+&  \left. {2 \over (m_{\phi_3}^2 + m_X^2)^2} + {1 \over m_X^2 (m_{\phi_3}^2 
+ m_X^2)} + {3 \over 8 m_X^4} \right],
\end{eqnarray}
where the sum over $h$, $E$ is for 3 families. The factor of 3 for $h$ is the 
number of colors, and the factor of 2 for $E$ is to include $N$ which has 
the same mass of $E$.  For the scalar final states $\phi_1 \bar{\phi}_1$, 
in addition to the exchange of $\phi_3$, there is also the direct 
$X_1 X_1 \phi_1 \bar{\phi}_1$ interaction.  Since $v_1 << v_2$, both $\phi_1^0$ 
and $\phi_1^-$ are physical particles to a very good approximation. Assuming 
as we do that $m_X$ is the smallest mass in Eq.~(34), we must have
\begin{equation}
\sigma v_{rel} < {41 g_N^4 \over 576 \pi m_X^2}.
\end{equation}
This puts an upper bound on $m_X$ for a given value of $\sigma v_{rel}$. 
Assuming $\sigma v_{rel} > 0.86$ pb from the requirement of relic abundance, 
and $g_N^2 (\simeq g_2^2) = 0.4$, we then obtain
\begin{equation}
m_X < 1.28~{\rm TeV}.
\end{equation}
In other words, whereas the scale of $SU(2)_N$ breaking is {\it a priori} 
unknown, the assumption of $X$ dark matter constrains it to be 
of order 1 TeV and be accessible to observation at the LHC.

We consider Eq.~(34) as a function of $m_X$ and $\delta = m_h/m_X - 1$, 
with all three $h$'s having the same mass.  We then consider the two 
extreme cases for the other contributions: one where all heavy masses are 
equal to $m_X$; and the other where all heavy masses (except $m_X$) are equal 
to the (arbitrary) value $2.5 m_X$ to ensure that no Yukawa or quartic 
coupling gets too 
large. In the $\delta-m_X$ plane, for a given value of 
$\sigma v_{rel}$, the region between these two lines is then the allowed 
parameter space for $m_X$ and $m_h$.  We show this in Fig.~2 for 
$\sigma v_{rel} = 0.91 \pm 0.05$ pb~\cite{pdg10}.
\begin{figure}[htb]
\begin{center}
\includegraphics[width=0.8\textwidth]{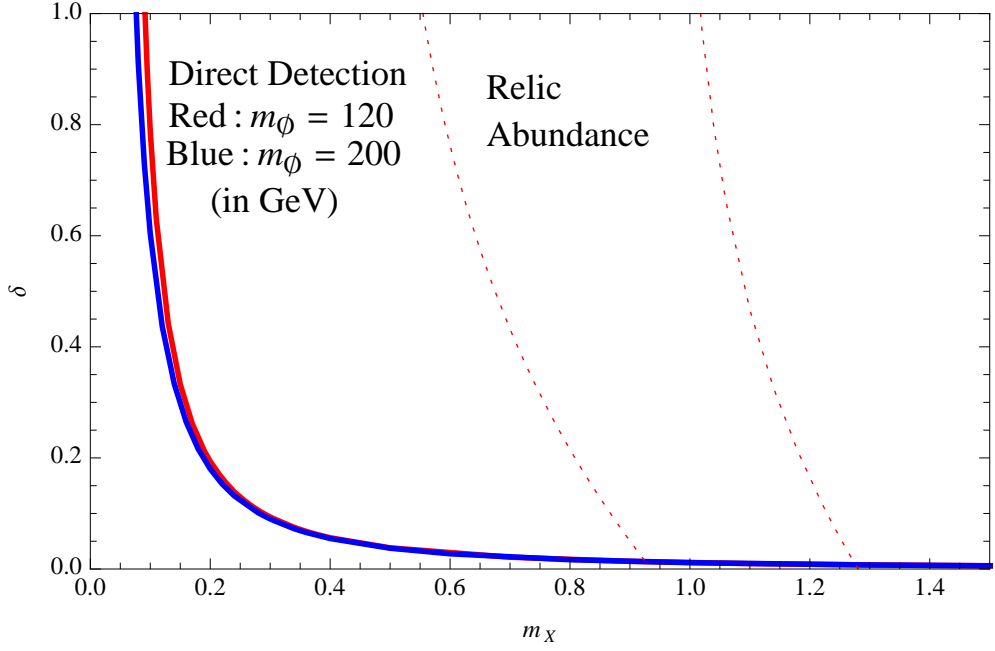}
\caption{Allowed region in $\delta=m_h/m_X - 1$ versus $m_X$ (in TeV) 
from relic abundance and from CDMS direct search.}
\end{center}
\end{figure}

\section{Direct dark matter search}

\begin{figure}[htb]
\begin{center}
\includegraphics[width=0.9\textwidth]{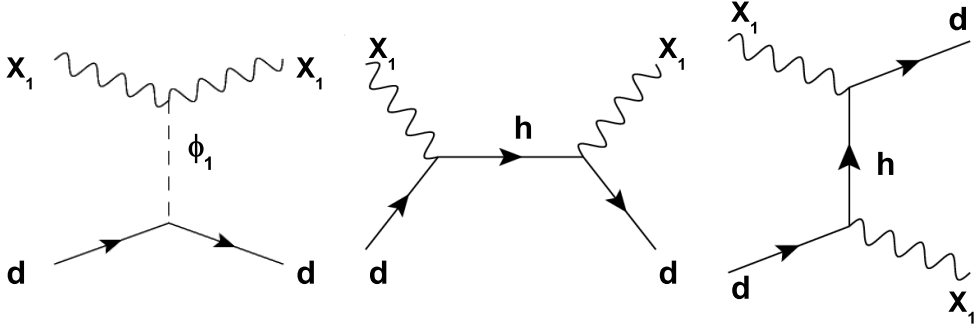}
\caption{Interactions of $X_1$ with quarks in direct-search 
experiments.}
\end{center}
\end{figure}
In Fig.~3 we show the tree-level diagrams for $X_1 d \to X_1 d$ through 
the direct-channel exchange of $h$ and the cross-channel exchange of 
$\phi_{1R}$.  Taking into account twist-2 operators and gluonic contributions 
calculated recently~\cite{hiny10} and assuming that $m(h_d) = m(h_s) = m(h_b) 
= m_h$, we find
\begin{eqnarray}
{f_p \over m_p} &=& 0.052 \left[ - {g_N^2 \over 4 m_\phi^2} - {g_N^2 \over 16} 
{m_h^2 \over (m_h^2-m_X^2)^2} \right] + {3 \over 4} (0.222) \left[ - 
{g_N^2 \over 4} {m_X^2 \over (m_h^2 - m_X^2)^2} \right] \nonumber \\ 
&-& (0.925) \left( (1.19){g_N^2 \over 54 m_\phi^2} + {g_N^2 \over 36} \left[ 
(1.19) {m_h^2 \over 6 (m_h^2-m_X^2)^2} + {1 \over 3(m_h^2-m_X^2)} \right] 
\right).
\end{eqnarray}
To obtain $f_n/m_n$, the numerical coefficients (0.052,0.222,0.925)  
in the above are replaced by (0.061,0.330,0.922).  The spin-independent 
elastic cross section for $X_1$ scattering off a nucleus of $Z$ protons 
and $A-Z$ neutrons normalized to one nucleon is then given by  
\begin{equation}
\sigma_0 = {1 \over \pi} \left( {m_N \over m_X} \right)^2 \left| 
{Z f_p + (A-Z) f_n \over A} \right|^2.
\end{equation}
Here we will use $^{73}Ge$ with $Z=32$ and $A-Z=41$ to compare against the 
recent CDMS result~\cite{cdms10}.  In the range $0.3 < m_X < 1.0$ TeV, 
the experimental upper bound is very well approximated by~\cite{klm10}
\begin{equation}
\sigma_0 < 2.2 \times 10^{-7} ~ {\rm pb}~ (m_X/ 1 ~{\rm TeV})^{0.86}.
\end{equation}
In Fig.~2 this appears as a solid line for $m_\phi = 120$ GeV, to the right 
(left) of which is allowed (forbidden) by the CDMS data.  If $m_\phi > 120$ 
GeV, this line will move slightly to the left.  It is seen that the 
relic-abundance constraint is indeed allowed, but direct search is still 
far away from testing this model.

\begin{figure}[htb]
\begin{center}
\includegraphics[width=0.55\textwidth,angle=-90]{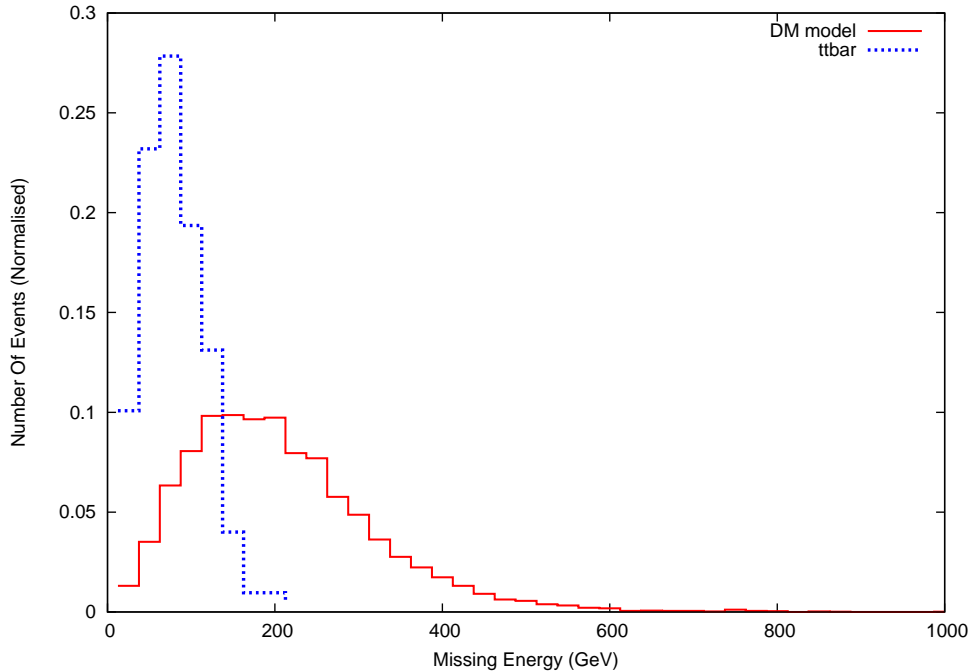}
\caption{Normalized signal and background distributions as functions of 
missing transverse energy.}
\end{center}
\end{figure}

\section{Collider phenomenology}

The dark-matter gauge boson $X_1$ may be produced at the Large Hadron Collider 
in association with the lightest exotic heavy quark $h$ through $d + 
{\rm gluon} \to h + X_1$.  Consider the following mass spectrum:
\begin{equation}
m_h > m_{X_2} > m_{E,N} > m_{X_1}.
\end{equation}
In that case, $h$ may decay into $X_1 d$ and $X_2 d$, then $X_2$ will decay 
into $E^+l^-,E^-l^+,\bar{N} \nu, N \bar{\nu}$, and $E^+ \to X_1 l^+$, 
$E^- \to X_1 l^-$, $\bar{N} \to X_1 \bar{\nu}$, $N \to X_1 \nu$.  This means 
that about 1/4 of the time, $p p \to h X_1$ will end up with one quark 
jet + missing energy + $l_i^+ l_j^-$ and $p p \to h \bar{h}$ will end up 
with two quark jets + missing energy + $l_i^+ l_j^-$.  Some of these 
two-lepton final states could involve different flavors because of mixing 
of families in the $SU(2)_N$ sector.  Note that $X_2 \to X_1 + 
{\rm virtual}~X_3 \to X_1 + d \bar{d}~(l^- l^+)$ is also possible, but 
very much suppressed if $m_{E,N} < m_{X_2}$.

In the following, we choose $m_{X_1} = 700$ GeV, $m_{E,N} = $ 735 GeV, 
$m_{X_2} = $ 770 GeV, and $m_h = 980$ GeV.  We find that at the LHC 
($E_{cm} = 14$ TeV), the cross section of $d X_1 X_1 l^- l^+$ production 
is 5.5 fb.  We show in Fig.~4 the distribution of this signal versus 
the expected standard-model background (dominated by $t \bar{t}$) 
as a function of missing transverse energy, 
using the cut $p_{T} > 20$ GeV for each lepton with $|\eta_l| < 2.5$, 
and $p_{T} > 50$ GeV for the one hadronic jet.  
We use {\tt CalcHEP}~\cite{calchep}
in combination with {\tt Pythia}~\cite{PYTHIA} in this calculation. 
We show in Table 1 that a cut on missing transverse energy of 200 GeV 
would eliminate the standard-model background which is dominated by 
$t \bar{t}$ events.
\begin{table}[htb]
\begin{center}
\begin{tabular}{|c|c||c|c||c|c|}
\hline
\multicolumn{6}{|c|}{Event rates for $\ell^+\ell^- + 1 jet + 
{E_T}\!\!\!\!\!\!\!/ \: \;$ with $p_{T_{\ell}} > 20$, $p_{T_{j}} > 50$} \\ 
\hline \hline
\multicolumn{2}{|c||}{${E_T}\!\!\!\!\!\!\!/$ $\quad>$ 100} & 
\multicolumn{2}{|c||}{${E_T}\!\!\!\!\!\!\!/$ $\quad>$ 200} & 
\multicolumn{2}{|c|}{${E_T}\!\!\!\!\!\!\!/$ $\quad>$ 300} \\ \hline
 Signal & Background  & Signal & Background   &  Signal & Background    \\ 
\hline  
 3.1   & 237        &   1.6  &   0          & 0.59      &    0       \\ 
\hline
\end{tabular}
\caption{Event rates (fb) for LHC with $E_{cm}=14$ TeV, using CTEQ6L parton 
distribution functions, and the average of final state particle masses 
as partonic $E_{cm}$.}
\end{center}
\end{table}

\section{Concluding remarks}

The (nonsupersymmetric) dark vector-gauge-boson model~\cite{dm11} is studied 
in some detail. Its complete particle content is delineated and analyzed, 
including the most general Higgs potential and its minimization.  The 
identification of the $X_1$ boson as a dark-matter candidate (to account 
for the observed relic abundance) constrains the $SU(2)_N$ breaking scale 
to be about 1 TeV.  We have updated the theoretical cross section for 
$X_1$ to interact in underground direct-search experiments.  The present 
CDMS bound is shown to be much below what is expected in this scenario. 
On the other hand, the prognosis for observing the consequences of this 
model at the LHC with $E_{cm} = 14$ TeV and integrated luminosity of 
10 fb$^{-1}$ is good, with an expected signal in our specific example 
of 16 events (dimuon + jet + missing energy) against negligible background 
for $m_{X_1} = 700$ GeV and $m_h = 980$ GeV.
 
\section*{Acknowledgements}

This work is supported in part by the US Department of Energy under Grant 
No.~DE-FG03-94ER40837, and by CONACYT-SNI (Mexico).  SB would like to thank 
Dr.~Ehsan Noruzifar for technical help and Dr.~AseshKrishna Datta for 
valuable comments on the numerical simulation.

\bibliographystyle{unsrt}

\end{document}